\documentclass[twocolumn,amsmath,amssymb,superscriptaddress,prl,floatfix]{revtex4}
\usepackage[]{graphicx}
\usepackage[usenames,dvipsnames]{color}
\usepackage{soul}
\usepackage{bm}

\usepackage{times}
\usepackage{amsfonts}
\usepackage{mathrsfs}
\usepackage{graphicx}
\usepackage{dcolumn}
\usepackage{bm}
\usepackage{color}

\usepackage[colorlinks,bookmarks=false,citecolor=blue,linkcolor=red,urlcolor=blue]{hyperref}
\usepackage{multirow}

\begin{document}

\title{Substrate-supported triplet superconductivity in Dirac semimetals}

\author{Domenico Di Sante$^\dag$}
\affiliation{Institut f\"{u}r Theoretische Physik und Astrophysik, Universit\"{a}t W\"{u}rzburg, Am Hubland Campus S\"{u}d, W\"{u}rzburg 97074, Germany}\email{domenico.disante@physik.uni-wuerzburg.de}
\thanks{$^\dag$ These authors equally contributed to the work.}

\author{Xianxin Wu$^\dag$}
\affiliation{Institut f\"{u}r Theoretische Physik und Astrophysik, Universit\"{a}t W\"{u}rzburg, Am Hubland Campus S\"{u}d, W\"{u}rzburg 97074, Germany}

\author{Mario Fink}
\affiliation{Institut f\"{u}r Theoretische Physik und Astrophysik, Universit\"{a}t W\"{u}rzburg, Am Hubland Campus S\"{u}d, W\"{u}rzburg 97074, Germany}

\author{Werner Hanke}
\affiliation{Institut f\"{u}r Theoretische Physik und Astrophysik, Universit\"{a}t W\"{u}rzburg, Am Hubland Campus S\"{u}d, W\"{u}rzburg 97074, Germany}

\author{Ronny Thomale}
\affiliation{Institut f\"{u}r Theoretische Physik und Astrophysik, Universit\"{a}t W\"{u}rzburg, Am Hubland Campus S\"{u}d, W\"{u}rzburg 97074, Germany}

\date{\today}

\maketitle

{\bf Stimulated by the success of graphene and its emerging Dirac
physics~\cite{RevGraphene}, the quest for versatile and tunable
electronic properties in atomically thin systems has led to the
discovery of various chemical classes of 2D compounds~\cite{2DXenes}. In
particular, honeycomb lattices of group-IV elements, such as silicene
and germanene, have been found
experimentally~\cite{SiliceneAg1,SiliceneZrB2,GermaneneAu,GermaneneAlN,
GermaneneMoS2}. Whether it is a necessity of synthesis or a desired
feature for application purposes, most 2D materials demand a supporting
substrate. In this work, we highlight the constructive impact of
substrates to enable the realization of exotic electronic quantum states
of matter, where the buckling emerges as the decisive material parameter
adjustable by the substrate. At the example of germanene deposited on
MoS$_2$, we find that the coupling between the monolayer and the
substrate, together with the buckled hexagonal geometry, conspire to
provide a highly suited scenario for unconventional triplet
superconductivity upon adatom-assisted doping.}

Material synthesis in two spatial dimensions is one of the rising
fields of contemporary condensed matter physics. Initiated by the
exfoliation and substrate-assisted growth of graphene~\cite{Geim2007},
complementary techniques such as refined sputtering and molecular beam
epitaxy are significantly broadening the scope of 2D material
classes, which by themselves are considered promising hosts for exotic
electronic quantum states of matter. This includes not only
topological quantum matter such as quantum spin Hall (QSH) insulators and
Chern insulators~\cite{QSHE2,Haldane}, but also unconventional superconductors,
which has recently climaxed in the discovery of superconductivity in
doped twisted bilayer graphene~\cite{TBG}. 

\begin{figure}[!t]
\centering
\includegraphics[width=0.8\columnwidth,angle=0,clip=true]{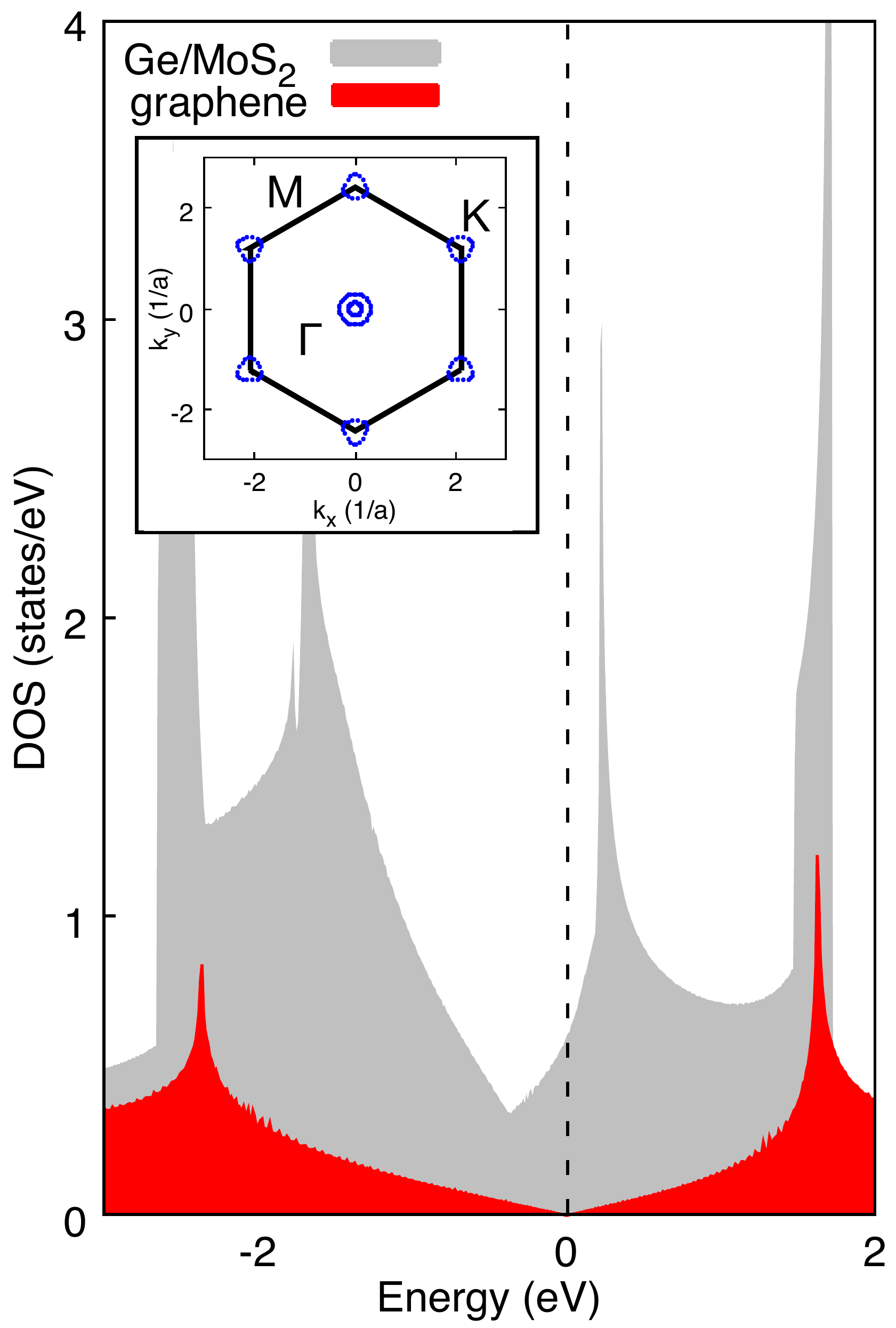}
\caption{{\bf Fermi surface and density of the states.} Comparison
between the DOS of graphene (red) and germanene/MoS$_2$ (grey). The
Fermi level, corresponding to the vertical dashed line, is set to
half-filling in both cases, and separates the hole doping (left) from
the electron doping (right) regimes. The inset shows the Fermi surface
of germanene/MoS$_2$ in the absence of doping.}
\label{fig1}
\end{figure}

For 2D superconductors, an overarching principle is to attempt to access
high density of states (DOS) at the Fermi level which constitutes a
promising setup for high critical temperatures. For graphene, at a
filling factor corresponding to $\sim$13\% of doping concentration, a
van Hove singularity (vHs) in the DOS was proposed to drive a
substantial enhancement of interaction effects~\cite{Dzyaloshinskii}.
One striking consequence is the predicted appearance of $d+id$-wave
superconductivity~\cite{Graphene3RG,GrapheneFRG,Wang2012}, which would
allow to enter the rich phenomenology of topological superconductors.
Several attempts have been made to dope graphene to the vH point by
chemical doping~\cite{DopedGraphene}. Notwithstanding the efforts, so
far no evidence for the observation of unconventional superconductivity
has been reported. This is presumably because of the added disorder
capping the large DOS at the vHs, which, in graphene, only shows up (see
Fig. \ref{fig1}) at energies, away from the Fermi level at half-filling,
comparable to the nearest-neighbor hopping parameter ($\sim$ 3 eV). As a
consequence, one guiding principle for improvement is to identify
alternative scenarios in which vHs filling can be achieved at lower
doping. 

Silicene and germanene as 2D-Xenes exhibit larger bond lengths than
graphene and prevent the atoms from forming strong $\pi$-bonds, yielding
a smaller nearest-neighbor hopping ($\sim$ 1 eV). The growth of such
systems requires proper substrates and templates. Aiming at QSH
insulator phases, 2D-Xenes may be suited because of their heavy
constituent atoms and the larger spin-orbit mediated topological band
gap~\cite{QSHE1,QSHE3,Reis287}. X-ene geometric reconstruction has been reported
for metallic substrates Ag, Au, Pt, Al and
Ir~\cite{SiliceneAg1,SiliceneZrB2,GermaneneAu}, as well as less
interacting substrates as MoS$_2$ and
AlN~\cite{GermaneneAlN,GermaneneMoS2}. Common reconstructed phases are
$\sqrt{3}\times\sqrt{3}$, $\sqrt{7}\times\sqrt{7}$, $2\times 2$, but
also the larger $3\times 3$ and $5\times 5$ setting~\cite{2DXenes}.
Bonding with the substrate and complex surface reconstructions render
the analysis of realistic systems a challenging task from experiment and
theory. Furthermore, the strong monolayer/substrate interaction often
avoids the QSH scenario in favour of, on a first view, an undesirable
metallic phase~\cite{GermaneneMoS2,Zhu2015}. In light of unconventional
superconductivity, however, the key insight of our work is that the
presence of a substrate can drastically modify the low-energy physics of
2D-Xenes in an advantageous manner, as to create a new fermiology
characterized by enlarged DOS already at pristine filling, and a vHs
accessible upon moderate doping. As the graphene-type fermiology is
fundamentally altered through the substrate, we find that
substrate-supported 2D electronic structures establish an intriguing
platform for unconventional Fermi surface instabilities in general, and
superconductivity in particular. By germanene on MoS$_2$, we identify an
electronic structure which promises to be preeminently suited for the
observation of $f$-wave superconductivity, a state which has so far
remained elusive in nature. 

\begin{figure*}[!t]
\centering
\includegraphics[width=\textwidth,angle=0,clip=true]{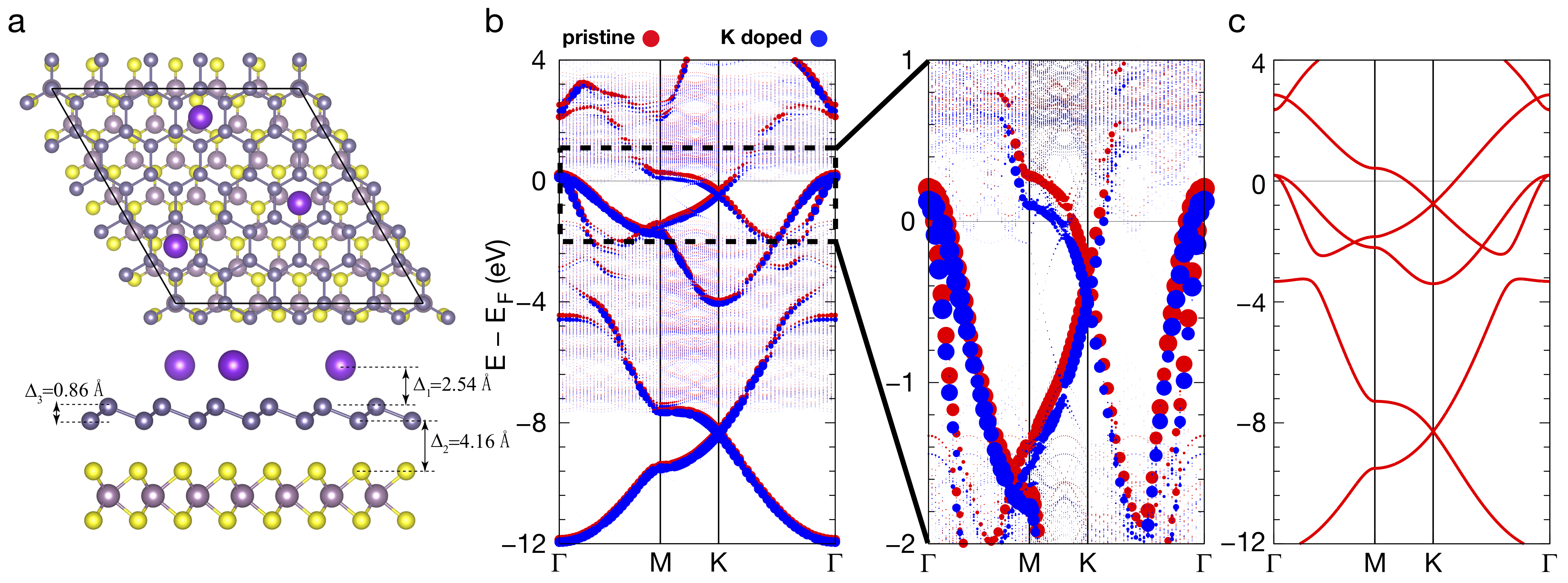}
\caption{{\bf First-principles calculations of realistic
germanene/MoS$_2$.} {\bf a} Structural model for $5\times 5$ germanene
on $6\times 6$ MoS$_2$, with the inclusion of potassium atoms to
simulate chemical doping. {\bf b} DFT bandstructure (wide energy range
view on the left and zoom around the Fermi level on the right) of the
superstructure in panel {\bf a} along the high-symmetry-lines of the
$1\times 1$ Brillouin zone. The red and blue circles highlight the
weights of the unfolded electronic states for the pristine and K-doped
system, respectively. {\bf c} Bandstructure of the nearest-neighbour
tight-binding (TB) Slater-Koster model with parameters listed in Table
\ref{tab1} and Hamiltonian given in the Methods section.}
\label{fig2}
\end{figure*}

\begin{table*}[t]
\centering
\caption{{\bf Structural and hopping parameters in the 8-bands NN
Slater-Koster tight-binding model.} $\Delta_3$ (\AA) is the buckling
parameter of the honeycomb lattice (see Fig. \ref{fig2}{\bf a}),
described also by the buckling angle $\theta$ ($^\circ$) between the
Ge-Ge bond and the $z$ direction normal to the surface. $\epsilon_{s}$,
$\epsilon_{p_{x,y}}$ and $\epsilon_{p_z}$ (eV) are the on-site energies
of the $s$, $p_x$, $p_y$ and $p_z$ orbitals, while V$_{ss\sigma}$,
V$_{sp\sigma}$, V$_{pp\sigma}$ and V$_{pp\pi}$ (eV) parametrize the
Slater-Koster transfer intergrals. DS, CDS and M in the last column are
acronym for Dirac Semimetal, Compensated Dirac Semimetal and Metallic
phases of the ground state, respectively. The bandstructures for
Ge/AlN/Ag(111) and Ge/Au(111) are shown in the Supplementary
Information.}
\label{tab1}
\begin{tabular}{p{2.5cm}p{3cm}p{1cm}p{1cm}p{1cm}p{1cm}p{1cm}p{1cm}p{1cm}p{1cm}p{1cm}p{1cm}}
\hline\hline
               & reconstruction                                    & $\Delta_3$     &  $\theta$             &   $\epsilon_{s}$       &   $\epsilon_{p_{x,y}}$       &   $\epsilon_{p_z}$       &  V$_{ss\sigma}$  &  V$_{sp\sigma}$  &  V$_{pp\sigma}$  &   V$_{pp\pi}$   &  phase \\
\hline
Ge/AlN/Ag(111) & $3\times 3$/$4\times 4$                           & 0.70           &  107.3                &   -5.44              &    2.76                    &    0.86                & -1.8             &  2.5             &  3.3             & -1.0     & DS      \\
Ge/MoS$_2$     & $5\times 5$/$6\times 6$                           & 0.86           &  111.4                &   -5.74              &    2.46                    &    0.56                & -2.0             &  2.5             &  3.3             & -1.2     & CDS     \\
Ge/Au(111)     & $\sqrt{3}\times\sqrt{3}$/$\sqrt{7}\times\sqrt{7}$ & 0.47           &  100.5                &   -6.24              &    1.96                    &    0.06                & -1.5             &  2.5             &  3.3             & -1.2     & M       \\
\hline
\end{tabular}
\end{table*}
 
The peculiar properties of Ge/MoS$_2$ already become visible from a
comparison of DOS against graphene (Fig. \ref{fig1}). Already at half
filling, Ge/MoS$_2$ exhibits a sizable carrier density. Even more
remarkably, however, the vHs, in particular the one on the
electron-doped side, is shifted closer to half filling as compared to
graphene. This observation is of general importance independent of the
type of Fermi surface instability we are interested in, since all
instability scales, from the viewpoint of weak coupling, are enhanced by
an enlarged DOS at the Fermi level. As a next step, we analyze the
detailed fermiology and general microscopic setting of Ge/MoS$_2$ in the
vicinity of vH filling.

Fig. \ref{fig2}{\bf a} shows the established structural model for
$5\times 5$ germanene on $6\times 6$ MoS$_2$~\cite{GermaneneMoS2}, while
in Table \ref{tab1} and in the Supplementary Information we also report
the cases of $3\times 3$ germanene on $4\times 4$
AlN/Ag(111)~\cite{GermaneneAlN} and $\sqrt{3}\times \sqrt{3}R(30^\circ)$
germanene on $\sqrt{7}\times \sqrt{7}R(19.1^\circ)$
Au(111)~\cite{GermaneneAu}. We computed the supercell's bandstructure
and unfolded it into the primitive Brillouin zone, obtaining the
unfolding weights (red symbols in Fig. \ref{fig2}{\bf b}, see Methods
for references). When grown on MoS$_2$, the electronic states of
germanene are weakly disturbed by the interaction with the substrate.
The significant compressive lateral strain on the honeycomb lattice
($\sim 5\%$), however, increases the buckling distortion up to
$0.86$\AA, and induces a crossing of the Fermi level around the $\Gamma$
point by hole-like bands (see also the Supplementary Information),
turning the system into a compensated Dirac semimetal phase with a
finite DOS at the Fermi level. We develop a realistic tight-binding
model able to reproduce the single particle bandstructure of
Ge/MoS$_2$~\cite{TBmodel}. In Table \ref{tab1}, we report our parameter
fit to the ab initio results within a simplified nearest-neighbour (NN)
approximation (Fig. \ref{fig2}{\bf c} for Ge/MoS$_2$).
Next-nearest-neighbor hoppings are delegated to the Supplementary
Information; as another step of refinement, instead of resorting to a
tight-binding fit, we also employ a full Wannier function based model,
which we have used for our Femi surface instability calculations. 

\begin{figure}[!b]
\centering
\includegraphics[width=\columnwidth,angle=0,clip=true]{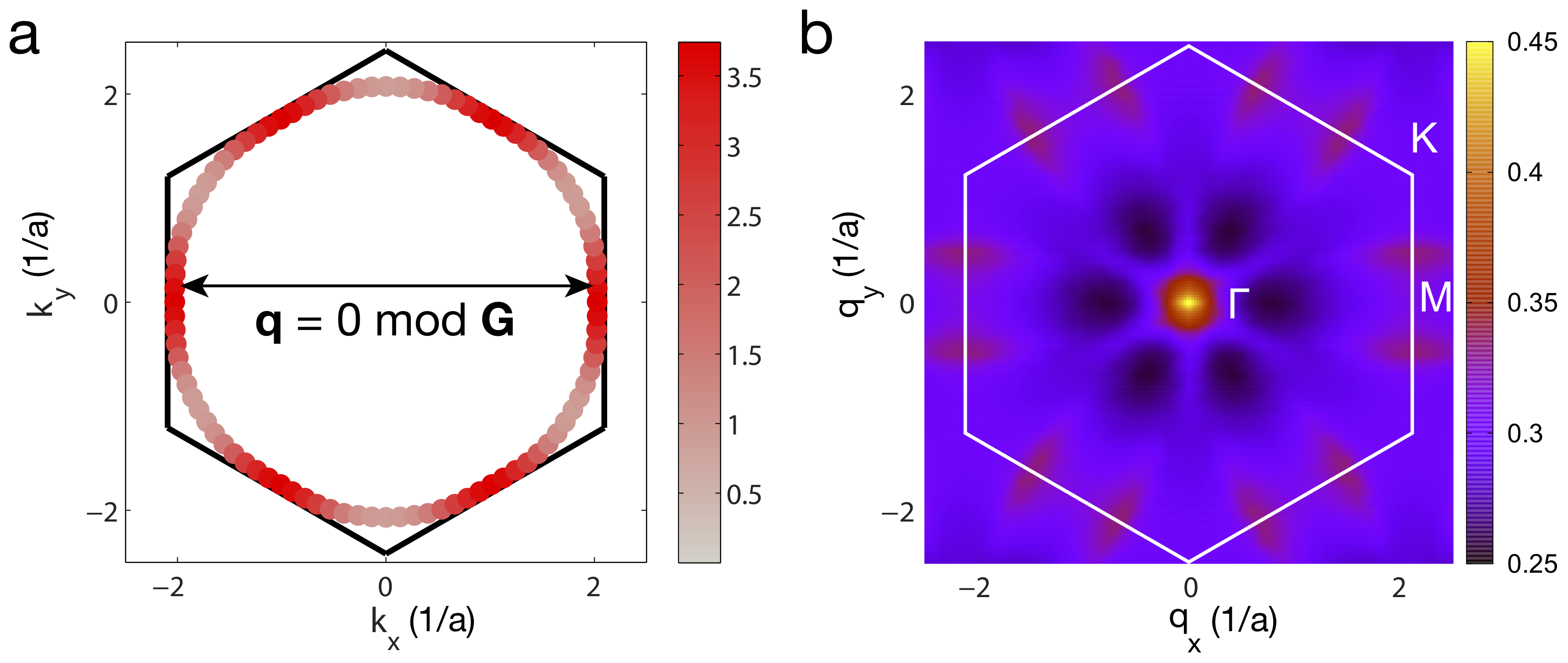}
\caption{{\bf Fermiology of germanene/MoS$_2$ at vHs point.} {\bf a}
Fermi surface and momentum distribution of the DOS ($\rho_{\bf{k}}\sim
1/|v_{\text F}(\bf{k})|$, with $v_{\text F}(\bf{k})$ the Fermi velocity)
at the vHs energy. The arrow highlights the ${\bf q}=0$ nesting vector
up to a reciprocal lattice vector. {\bf b} Momentum distribution of the
RPA bare susceptibility (see Methods). The color bars are in eV\AA~and
eV$^{-1}$ units, respectively.}
\label{figvf}
\end{figure}

In graphene, 0.5 electron doping per unit cell is needed to reach the
vHs point. At the present stage of experimental capabilities, such a
high electron doping is unavoidably accompanied by detrimental disorder
effects.  In Ge/MoS$_2$, as the vHs point is energetically closer to the
Fermi level, the vHs can be reached upon doping of $\sim 0.2$ electrons
per unit cell, i.e., only 40$\%$ of the doping value needed for
graphene. In Fig. \ref{fig2}{\bf b}, we show the bandstructure of
Ge/MoS$_2$ upon doping by 3 alkali atoms per unit cell. This doping
shifts the vHs close to the Fermi level (without much affecting the
states around the $\Gamma$ point) by providing 0.12 electrons, such that
the vHs now is only 0.1 eV above the Fermi level.

The Fermi surface of Ge/MoS$_2$ doped to the vHs is shown in
Fig.\ref{figvf}{\bf a}. It is almost circular and rather flat along the
$M-K$ line, which is in sharp contrast to the expected hexagonal shape
for vH-doped graphene. In graphene, the nesting between opposite edges
of the hexagonal Fermi surface promotes strong antiferromagnetic
fluctuations around the $M$ point, which in turn drive the $d+id$
pairing states~\cite{Graphene3RG,GrapheneFRG,Wang2012}. In Ge/MoS$_2$,
this nesting is absent due to the circular Fermi surface, and the
dominant nesting (denoted by the arrow) promotes ferromagnetic
fluctuations. This is evident from the intense ${\bf q}=0$ peak in the
momentum space distribution of the bare susceptibility shown in
Fig.\ref{figvf}{\bf b}. Furthermore, the Fermi velocities of graphene
and Ge/MoS$_2$ are different. For the former, the minimum of the Fermi
velocity is localized around the $M$ point, leading to a peaked DOS in
its vicinity. For the latter, a high DOS is extended over a large
fraction of the Fermi surface (see Fig.\ref{figvf}{\bf a}), and any
minimal reduction to the $M$ points is no longer
valid~\cite{Graphene3RG}.

We adopt the functional renormalization group (FRG) approach to study
the Fermi surface instabilities, by starting from the bare many-body
interaction and integrating out the high-energy degrees of
freedom~\cite{MetznerRMP,PlattReview}. The renormalized interaction
$V_\Lambda({\bf k}_1,{\bf k}_2,{\bf k}_3,{\bf k}_4)$ described by the
4-point function (4PF) diverges in some channels as the cutoff $\Lambda$
approaches the Fermi surface, marking the onset of a leading instability
(see Methods). The parameterization of germanene on MoS$_2$ (bandwidth W
$\sim 20$ eV) serves as the non-renormalized limit for our FRG study.
The interaction Hamiltonian we consider contains intra- and
inter-orbital repulsion $U$ and $U'$, as well as Hund's rule coupling
$J$ and pair-hopping $J'$ (see Methods). For simplicity, in the absence
of ab-initio estimates of the interaction parameters, we choose the
ansatz $U=U'+2J$, $U=2U'$, and $J = J'$, tuning the global scale such
that the resulting maximum strength of the initial vertex function
$V_{\Lambda=W}({\bf k}_1,{\bf k}_2,{\bf k}_3,{\bf k}_4)$ for momenta on
the Fermi surface is still located in the weak to intermediate coupling
regime. We use the same Hamiltonian within a multiband random phase
approximation (RPA) fluctuation exchange approximation
scheme~\cite{Bickers1989} in order to provide an independent validation
of our FRG results. Note that spin-orbit coupling (SOC) has not been
taken into account, which promotes a more efficient implementation. This
assumption is justified for germanene: as a recent theoretical study
demonstrates, in Ge/MoS$_2$ the layer/substrate interaction sensibly
reduces the influence of SOC compared to the freestanding
case~\cite{Menno}.

\begin{figure*}[!t]
\centering
\includegraphics[width=\textwidth,angle=0,clip=true]{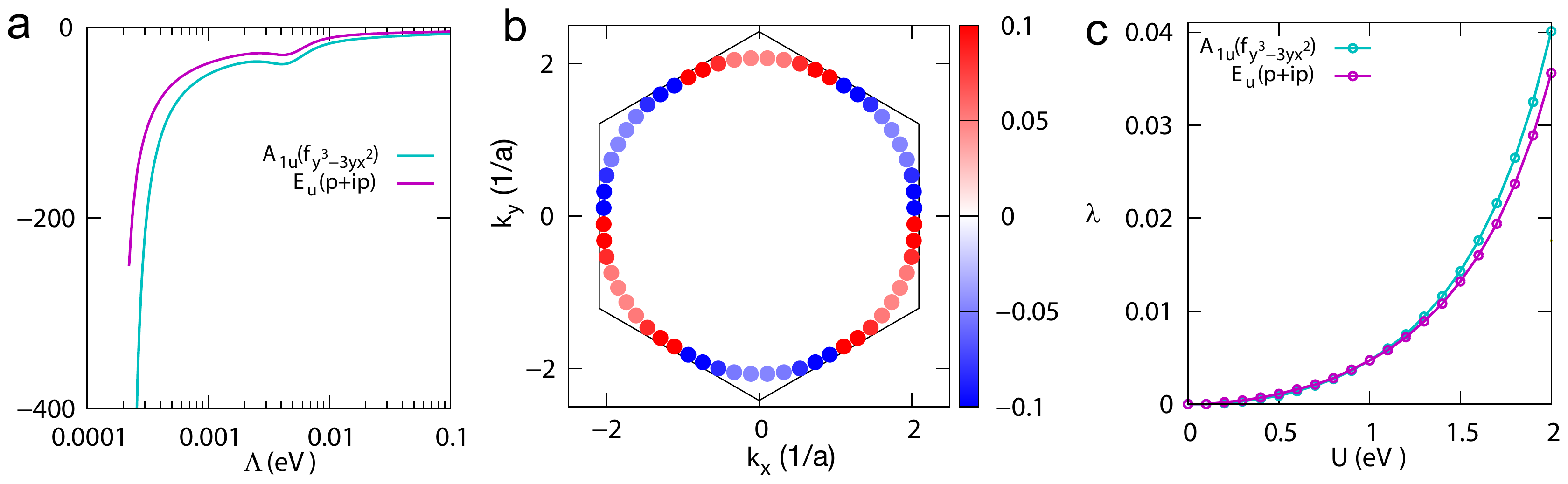}
\caption{{\bf Superconducting instability.} {\bf a} Typical FRG flow of
the Fermi surface instabilities when the chemical potential is set
slightly below the vHs as a function of the infrared cutoff $\Lambda$
(max$|V_{\Lambda=W}({\bf k}_1,{\bf k}_2,{\bf k}_3,{\bf k}_4)| = 2.0$
eV). The superconducting channels of the mean-field decomposition are
labelled according the irreducible representations of $D_{3d}$ they
transform to. {\bf b} Modulation of the superconducting gap of the
leading $f-$wave instability along the Fermi surface (arb. units). {\bf
c} Typical RPA results at the vHs as a function of the intra-orbital
repulsion $U$ at fixed $J/U=0.2$ ratio. $\lambda$s are the eigenvalues
of the RPA pairing vertex (see Methods).}
\label{fig3}
\end{figure*}

Near the vHs, we find a prominent superconducting instability in the
spin triplet sector, since the renormalized vertex $V_\Lambda$ diverges
in this channel as $\Lambda$ approaches in the FRG flow the Fermi level
(see Fig. \ref{fig3}{\bf a}). The leading pairing state transforms
according to the $A_{1u}$ irreducible representation of $D_{3d}$, i.e.,
the point group symmetry of the buckled geometry of germanene. Imposing
a mean field decoupling in the diverging vertex channel~\cite{Rohe2007},
we find that the gap function $\Delta_{\bf{k}}$ associated to the
leading instability shows a $f_{y(y^2-3x^2)}$ profile, where the gap
function changes sign every 60$^\circ$ rotation and has line nodes along
the $k_y=0$, $k_y=\pm\sqrt{3}k_x$ lines (see Fig. \ref{fig3}{\bf b}).
The FRG results are validated by RPA fluctuation exchange calculations,
the results of which we show in Fig. \ref{fig3}{\bf c}, where we also
find the dominant pairing state to be located in the $A_{1u}$ lattice
group representation (see Methods section for details). As the Fermi
pocket is rather circular and does not exhibit a particularly peaked
momentum structure of the DOS at the Fermi level, the ${\bf q}=0$
particle-hole fluctuation channel is dominant. This naturally promotes
the tendendy towards triplet pairing, where all sub-channels satisfy the
condition that the gap function connected by the nesting vector ${\bf
q}$ (see Fig.\ref{figvf}{\bf a}) must have the same sign. This is a
recurrent motif from other theoretical proposals for $f$-wave
superconductivity, such as employing the sublattice interference in a
kagome metal~\cite{KieselKagome}. 

Within the triplet channel, microscopic details such as the hexagonal
symmetry then yield a preference of the $f$-wave state over other
candidates states such as the $p$-wave state, which is the subleading
instability both at the FRG and RPA level. We also note that the
competition between the $f-$ and $p-$wave instability can to some extent
be tuned by varying the $J/U$ ratio in the interaction Hamiltonian. This
is an interesting perspective if we assume that the Hund's coupling can
be tailored by substrate engineering. The agreement between FRG and RPA
significantly supports the prediction of $f$-wave triplet
superconductivity in Ge/MoS$_2$. This is because it might sometimes
occur that the relevance of ferromagnetic fluctuations is overestimated.
At the instance of LiFeAs, early RPA studies had predicted $p$-wave
superconductivity~\cite{Brydon2011}, whereas FRG found a dominant
extended $s$-wave which agreed with the finally converging picture from
experimental evidence~\cite{Platt2011}.

Our combined ab initio, RPA, and FRG analysis points out that the
recently synthesized 2D-Xene  Ge/MoS$_2$ is a promising platform for
studying unconventional Fermi surface instabilities, in light of triplet
superconductivity. In order to reach the scenario outlined here, further
steps of experimental refinement suggest themselves to be followed up
on. First, there appears to be an electronic level mismatch between the
current Ge/MoS$_2$ in experiment and the ab initio simulations, possibly
due to non-saturated defects \cite{GermaneneMoS2}. Second, to avoid
detrimental disorder effects from chemical doping, one might want to
pursue electrostatic doping from gating methods. Recent experiments have
achieved a doping in MoS$_2$ of $\sim 1.2\times10^{14}$cm$^{-2}$ carrier
density~\cite{Ye1193}, which, if transferred to germanene, would
correspond to $\sim 0.08$ electrons. At this doping level, our
calculations already suggest a propensity towards a
$f_{x(x^2-3y^2)}-$wave instability, even though higher doping would
still be desirable. From a broader perspective, this is only the
beginning to employ {\it substrate engineering} towards accomplishing
exotic Fermi surface instabilities. As significant progress has already
been made at the frontier of substrate-assisted topological
insulators~\cite{Reis287}, we hope that our work will stimulate similar
efforts for unconventional superconductivity in layer/substrate
heterostructures.\\

{\bf Methods}\\

{\bf The model.} The eight-band tight binding model with a simple
nearest neighbor hopping parameterization for germanene is given by
\begin{eqnarray}
H_{0}=\sum_{\alpha\beta}\sum_{\mu\nu\sigma}h^{\alpha\beta}_{\mu\nu}(\mathbf{k})c^\dag_{\alpha\mu\sigma}(\bm{k})c_{\beta\nu\sigma}(\bm{k}) \nonumber,
\end{eqnarray}
where $\alpha/\beta=A,B$ is the sublattice index and $\mu/\nu=1,2,3,4$
represent the $s$, $p_z$, $p_x$ and $p_y$ orbitals, respectively.
$c^\dag_{\alpha\mu\sigma}$ creates a spin $\sigma$ electron in $\mu$
orbital on $\alpha$ sublattice with momentum $\bm{k}$. The matrix
elements $h^{\alpha\beta}_{\mu\nu}(\mathbf{k})$ are given by,
\begin{eqnarray}
h^{AA}(\mathbf{k})&=&h^{BB}(\mathbf{k})=diag(\epsilon_s,\epsilon_{p_z},\epsilon_{p_x},\epsilon_{p_y}) \nonumber\\
h^{AB}_{11}(\mathbf{k})&=&r^{11}(e^{ix}+2e^{-\frac{ix}{2}}\cos y) \nonumber\\
h^{AB}_{22}(\mathbf{k})&=&t^{11}(e^{ix}+2e^{-\frac{ix}{2}}\cos y) \nonumber\\
h^{AB}_{12/21}(\mathbf{k})&=&\pm s^{11}(e^{ix}+2e^{-\frac{ix}{2}}\cos y) \nonumber \\
h^{AB}_{13/31}(\mathbf{k})&=&\pm(s^{12}_1e^{ix}-s^{12}_1e^{-\frac{ix}{2}}\cos y)   \nonumber\\
h^{AB}_{14/41}(\mathbf{k})&=&\pm\sqrt{3}is^{12}_1e^{-\frac{ix}{2}}\sin y \nonumber\\
h^{AB}_{23/32}(\mathbf{k})&=&t^{12}_1e^{ix}-t^{12}_1e^{-\frac{ix}{2}}\cos y  \nonumber\\
h^{AB}_{24/42}(\mathbf{k})&=&\sqrt{3}it^{12}_1e^{-\frac{ix}{2}}\sin y \nonumber\\
h^{AB}_{33}(\mathbf{k})&=&t^{22}_{11}e^{ix}+\frac{1}{2}(t^{22}_{11}+3t^{22}_{22})e^{-\frac{ix}{2}}\cos y \nonumber\\
h^{AB}_{34/43}(\mathbf{k})&=& -\frac{\sqrt{3}i}{2}(t^{22}_{11}-t^{22}_{22})e^{-\frac{ix}{2}}\sin y \nonumber\\
h^{AB}_{44}(\mathbf{k})&=&t^{22}_{22}e^{ix}+\frac{1}{2}(3t^{22}_{11}+t^{22}_{22})e^{-\frac{ix}{2}}\cos y \nonumber
\end{eqnarray}
where $x = k_x a_0/\sqrt{3}$, $y = k_y a_0/2$,  $a_0$ is the in-plane
lattice constant, $r$ represents hopping between $s-s$ orbitals, $t$
represents hopping between $p-p$ orbitals and $s$ represents hopping
between $s-p$ orbitals. The hopping parameters in the model are,
\begin{eqnarray}
&& r^{11}=V_{ss\sigma} \nonumber\\
&& t^{11}= V_{pp\sigma}\cos^2\theta+ V_{pp\pi}\sin^2\theta \nonumber\\
&& s^{12}_1= V_{sp\sigma}\sin\theta \nonumber\\
&& s^{11}_1= V_{sp\sigma}\cos\theta \nonumber\\
&& t^{12}_1=- (V_{pp\pi}-V_{pp\sigma})\cos\theta \sin\theta \nonumber\\
&& t^{22}_{11}= V_{pp\sigma}\sin^2\theta+ V_{pp\pi}\cos^2\theta \nonumber \\
&& t^{22}_{22}=V_{pp\pi} \nonumber
\end{eqnarray}
with $\theta$ and Slater-Koster parameters $V$ listed in Table I of the
main text. A more elaborated model which includes next-nearest-neighbour
hopping terms is given in the Supplementary Information. For the actual
FRG and RPA calculations shown in the main text we used an ab initio
Hamiltonian based on Wannier Functions, which including long range
hopping terms, gives the best description of the DFT bandstructure.

The interaction part of the Hamiltonian for the multi-orbital system
considered here is
\begin{eqnarray}
H_{int}&=&U\sum_{i\alpha}n_{i\alpha\uparrow}n_{i\alpha\downarrow}+U'\sum_{i,\alpha<\beta}n_{i\alpha}n_{i\beta} + \nonumber \\
&&J\sum_{i,\alpha<\beta,\sigma\sigma'}c^{\dag}_{i\alpha\sigma}c^{\dag}_{i\beta\sigma'}c_{i\alpha\sigma'}c_{i\beta\sigma} + \nonumber \\
&&J'\sum_{i,\alpha\neq\beta}c^{\dag}_{i\alpha\uparrow}c^{\dag}_{i\alpha\downarrow}c_{i\beta\downarrow}c_{i\beta\uparrow}
\label{interaction}
\end{eqnarray}
where $n_{i\alpha}=n_{\alpha\uparrow}+n_{\alpha\downarrow}$. $U$, $U'$,
$J$ and $J'$ parametrize the intra- and inter-orbital repulsion, as well
as the Hund's rule and pair-hopping terms, respectively.\\

{\bf DFT calculations.} We employed first-principles calculations based
on the density functional theory as implemented in the Vienna ab initio
simulation package (VASP)\cite{VASP1}, within the
projector-augmented-plane-wave (PAW) method\cite{VASP2,PAW}. The
generalized gradient approximation as parametrized by the PBE-GGA
functional for the exchange-correlation potential was used\cite{PBE}, by
expanding the Kohn-Sham wavefunctions into plane-waves up to an energy
cut-off of 600 eV, and we sampled the Brillouin zone on an
$8\times8\times1$ regular mesh. The large structural reconstruction
enforces a folding of the electronic states into the supercell Brillouin
zones, which map onto the primitive $1\times1$ Brillouin zone. It is
usually simpler to achieve a transparent physical description in the
latter setting, where the unfolded bandstructure readily compares with
the freestanding models when the symmetry breaking induced by the
reconstruction is weak. The unfolding procedure we adopt in this work
follows the lines described in Refs\cite{WeiKu,vaspunfold}.\\

{\bf FRG calculations.} The basic idea of the functional Renormalization
Group (FRG) method is similar to other RG concepts in that the
"high-energy" degrees of freedom are thinned out while leaving the
low-energy physics invariant. In this way, FRG has become a much-used
general concept to derive effective theories, e.g. at long length scales
or for a low-energy window. For weakly to intermediately coupled Fermion
systems, one is mainly interested in the effective interactions near the
Fermi surface (E$_F$), as they contain relevant information about
possibly symmetry-broken (magnetic, superconducting, etc) and other
non-Fermi-liquid ground states. Therefore, by systematically integrating
out degrees of freedom, one can access competing orders at low-energy or
temperature in the phase diagram\cite{MetznerRMP,PlattReview}.

The RG approaches to interacting Fermions are less biased than
diagrammatic summations in a particular channel, as competing
fluctuations (magnetic, superconducting, screening, vertex corrections)
are included on equal footing. This differs, in particular, from the
random-phase-approximation (RPA) which, considering superconductivity,
takes right from the outset a magnetically spin-fluctuation driven
pairing interaction.

To compute the effective interactions near E$_F$, one typically uses the
RG flow equations for the effective action, or, one-particle irreducible
vertex functions\cite{MetznerRMP,PlattReview}. These schemes are named
functional RG, as they aim at keeping as much as possible the wave
vector dependence of the two-particle interaction V$_{\Lambda}({\bf
k},{\bf p})$, where $\Lambda$ denoted the RG flow parameter.

In the FRG one starts from the bare many-body interaction. The pairing
is then "dynamically" generated by integrating out the high-energy
degrees of freedom. For a given instability, characterized by the order
parameter $\hat{\text O}_{\bf k}$, the 4-point function (4PF)
V$_{\Lambda}({\bf k},{\bf p})$, in the particular ordering channel, can
be written in shorthand notation as $\sum_{{\bf k},{\bf p}}{\text
V}_{\Lambda}({\bf k},{\bf p})[\hat{\text O}_{\bf k},\hat{\text O}_{\bf
p}]$\cite{MetznerRMP,PlattReview,PlattFRG}. Accordingly, the 4PF
V$_{\Lambda}({\bf k}_i-{\bf k}_j,{\bf p}_i-{\bf p}_j)$ in the Cooper
channel can be decomposed into different contributions:

\begin{eqnarray}
\label{eq:2}
{\text V}_{\Lambda}^{\text{SC}}({\bf k},{\bf p}) = \sum_i c_i^{\text{SC}}(\Lambda)f_i^{\text{SC}*}({\bf k})f_i^{\text{SC}}({\bf p})
\end{eqnarray}
where $i$ is a symmetry decomposition index. The leading instabilities
of that channel then corresponds to an eigenvalue
$c_i^{\text{SC}}(\Lambda)$ first diverging under the flow of $\Lambda$.
$f_i^{\text{SC}}({\bf k})$ is the superconducting form factor of the
pairing mode $i$, which tells us about the superconducting pairing
symmetry and, hence, gap structure associated with it. In the FRG
scheme, from the final Cooper channel 4PF, this quantity is computed
along the discretized Fermi surfaces (as shown in Fig. \ref{fig3}{\bf
b}), and the leading instabilities are plotted in Fig. \ref{fig3}{\bf a}.

{\bf RPA fluctuation exchange calculations.} We adopt the standard multi-orbital RPA
approach\cite{Bickers1989,Kemper2010,XianxinWu1,XianxinWu2,XianxinWu3},
with parameter notations given in Ref.~\onlinecite{Kemper2010}. The bare
susceptibility is define as
\begin{eqnarray}
\chi^{0}_{l_1l_2l_3l_4}(\bm{q},\tau)&=&\frac{1}{N}\sum_{\bm{k}\bm{k}'}\langle T_{\tau} c^{\dag}_{l_3\sigma}(\bm{k}+\bm{q},\tau)c_{l_4\sigma}(\bm{k},\tau) \nonumber \\
&&c^{\dag}_{l_2\sigma}(\bm{k}'-\bm{q},0)c_{l_1\sigma}(\bm{k}',0) \rangle_0 .
\end{eqnarray}
where $l_i$ is the orbital indices. The bare susceptibility in
momentum-frequency space is then given by
\begin{eqnarray}
&&\chi^0_{l_1l_2l_3l_4}(\bm{q},i\omega_n)\!\!=\!\!-\frac{1}{N}\!\!\sum_{{\bm k}\mu\nu}a^{l_4}_\mu(\bm{k})a^{l_2*}_{\mu}(\bm{k}) a^{l_1}_\nu(\bm{k}+\bm{q})\times \nonumber \\
&&a^{l_3*}_{\nu}(\bm{k}+\bm{q})\frac{n_F(E_{\mu}(\bm{k}))-n_F(E_{\nu}(\bm{k}+\bm{q}))}{i\omega_n+E_{\mu}(\bm{k})-E_{\nu}(\bm{k}+\bm{q})},
\end{eqnarray}
where $\mu/\nu$ is the band index, $n_F(\epsilon)$ is the Fermi
distribution function, $a^{l_4}_\mu(\bm{k})$ is the $l_4$-th component
of the eigenvector for band $\mu$ resulting from the diagonalization of
the single particle Hamiltonian $H_0$ and $E_{\mu}(\bf{k})$ is the
corresponding eigenvalue. The interacting spin susceptibility and charge
susceptibility at the RPA level are given by
\begin{eqnarray}
\chi^{RPA}_1(\bm{q})&=&[1-\chi_0(\bm{q})U^s]^{-1}\chi_0(\bm{q}),\\
\chi^{RPA}_0(\bm{q})&=&[1+\chi_0(\bm{q})U^c]^{-1}\chi_0(\bm{q}),
\label{RPA1}
\end{eqnarray}
where $U^s$, $U^c$ are the interaction matrices:
\begin{eqnarray}
U^s_{l_1l_2l_3l_4}&=&
\begin{cases}
U   & l_1=l_2=l_3=l_4,\\
U'   & l_1=l_3\neq l_2=l_4,\\
J   & l_1=l_2\neq l_3=l_4,\\
J'   & l_1=l_4\neq l_2=l_3,\\
\end{cases}\\
\label{EQ:Us}
U^c_{l_1l_2l_3l_4}&=&
\begin{cases}
U   & l_1=l_2=l_3=l_4,\\
-U'+2J   & l_1=l_3\neq l_2=l_4,\\
2U'-J   & l_1=l_2\neq l_3=l_4,\\
J'   & l_1=l_4\neq l_2=l_3,\\
\end{cases}
\label{EQ:Uc}
\end{eqnarray}
The effective interaction within the RPA approximation is thus
\begin{eqnarray} V_{eff}=\sum_{ij,\textbf{k}\textbf{k}'}\Gamma_{ij}(\textbf{k},\textbf{k}')c^{\dag}_{i\textbf{k}\uparrow}c^{\dag}_{i-\textbf{k}\downarrow}c_{j-\textbf{k}'\downarrow}c_{j\textbf{k}'\uparrow}
\end{eqnarray}
where the momenta $\textbf{k}$ and $\textbf{k}'$ are restricted to 
different FS $C_i$ with $\textbf{k}\in C_i$ and $\textbf{k}'\in C_j$ and
$\Gamma_{ij}(\textbf{k},\textbf{k}')$ is the pairing scattering vertex
in the singlet channel\cite{Kemper2010}. The pairing vertex is,
\begin{eqnarray}
&&\Gamma_{ij}(\textbf{k},\textbf{k}')=Re[\sum_{l_1 l_2 l_3 l4}a^{l_2,*}_{v_i}(\textbf{k}) a^{l_3,*}_{v_i}(-\textbf{k})\times \nonumber \\
&&\Gamma_{l_1 l_2 l_3 l_4}(\textbf{k},\textbf{k}',\omega=0) a^{l_1}_{v_j}(\textbf{k}') a^{l_4}_{v_j}(-\textbf{k}')],
\end{eqnarray}
where $a^{l}_{v}$(orbital index $l$ and band index $v$) is the component
of the eigenvectors from the diagonalization of the tight binding
Hamiltonian. The orbital vertex function $\Gamma_{l_1 l_2 l_3 l_4}$ for
the singlet channel and triplet channel in the fluctuation exchange
formulation\cite{Bickers1989,Kemper2010} are given by,
\begin{eqnarray}
\Gamma^S_{l_1 l_2 l_3 l_4}(\textbf{k},\textbf{k}',\omega)&=&[\frac{3}{2}\bar{U}^s \chi^{RPA}_1(\textbf{k}-\textbf{k}',\omega)\bar{U}^s+\frac{1}{2}\bar{U}^s - \nonumber \\
&&\frac{1}{2}\bar{U}^c\chi^{RPA}_0(\textbf{k}-\textbf{k}',\omega)\bar{U}^c+\frac{1}{2}\bar{U}^c]_{l_1 l_2 l_3 l_4}, \nonumber \\
\Gamma^T_{l_1 l_2 l_3 l_4}(\textbf{k},\textbf{k}',\omega)&=&[-\frac{1}{2}\bar{U}^s \chi^{RPA}_1(\textbf{k}-\textbf{k}',\omega)\bar{U}^s+\frac{1}{2}\bar{U}^s -\nonumber \\
&&\frac{1}{2}\bar{U}^c\chi^{RPA}_0(\textbf{k}-\textbf{k}',\omega)\bar{U}^c+\frac{1}{2}\bar{U}^c]_{l_1 l_2 l_3 l_4}. \nonumber
\end{eqnarray}
The $\chi^{RPA}_0$ describes the charge fluctuation contribution and the
$\chi^{RPA}_1$ the spin fluctuation contribution. For a given gap
function $g(\textbf{k})$, the pairing strength functional is
\begin{eqnarray}
&&\lambda[g(\textbf{k})]= \nonumber \\
&-&\frac{\sum_{ij}\oint_{C_i} \frac{dk_{\|}}{v_F(\textbf{k})} \oint_{C_j} \frac{dk'_{\|}}{v_F(\textbf{k}')} g(\textbf{k})\Gamma_{ij}(\textbf{k},\textbf{k}') g(\textbf{k}')} {4\pi^2\sum_i\oint_{C_i} \frac{dk_{\|}}{v_F(\textbf{k})} [g(\textbf{k})]^2 },
\label{strength}
\end{eqnarray}
where $v_F(\textbf{k})=|\triangledown_{\textbf{k}}E_i(\textbf{k})|$ is
the Fermi velocity on a given fermi surface sheet $C_i$. From the
stationary condition we find the following eigenvalue problem
\begin{eqnarray}
-\sum_{j} \oint_{C_j} \frac{dk'_{\|}}{4\pi^2v_F(\textbf{k}')} \Gamma_{ij}(\textbf{k},\textbf{k}') g_{\alpha}(\textbf{k}')=\lambda_{\alpha}g_{\alpha}(\textbf{k}),
\label{strength1}
\end{eqnarray}
where the interaction $\Gamma_{ij}$ is the symmetric (antisymmetric)
part of the full interaction in singlet (triplet) channel. The leading
eigenfunction $g_{\alpha}(\bf{k})$ and eigenvalue $\lambda_{\alpha}$
(shown in Fig. \ref{fig3}{\bf c}) are obtained from the above equation.
The gap function has the symmetry of one of the irreducible
representations for the corresponding point group.

The interaction parameters that are used in the RPA analysis, in order
to avoid the magnetic instability, are systematically smaller than the
FRG counterparts. Moreover, in the FRG, the bare interaction is
renormalized to a smaller value as a consequence of the coupling between
the particle-hole and particle-particle channels. This screening, absent
in the RPA, justifies the necessity to use smaller bare interactions.\\

{\bf Acknowledgements.} This work was supported by the DFG through
SFB1170 "ToCoTronics" and by ERC-StG-336012-Thomale-TOPOLECTRICS. We
gratefully acknowledge the Gauss Centre for Supercomputing e.V.
(www.gauss-centre.eu) for funding this project by providing computing
time on the GCS Supercomputer SuperMUC at Leibniz Supercomputing Centre
(www.lrz.de).\\

{\bf Author contributions.} R.T., W.H and D.D.S conceived the project.
X.W. and D.D.S performed the DFT calculations while D.D.S, X.W. and M.F.
performed the FRG calculations and many-body analysis. X.W. performed
the RPA calculations. D.D.S, W.H. and R.T. wrote the manuscript and all
the authors equally contributed to the scientific discussion. X.W. and
D.D.S. equally contributed to the work.\\

{\bf Competing interests.} The authors declare no competing interests.\\

{\bf Additional information.} Correspondence and requests for materials
should be addressed to D.D.S.\\

\bibliography{biblio}

\newpage


\onecolumngrid
\begin{center}
\textbf{\large Supplementary Information for "Substrate-supported triplet superconductivity in Dirac semimetals"}
\end{center}
\setcounter{equation}{0}
\setcounter{figure}{0}
\setcounter{table}{0}
\setcounter{page}{1}
\makeatletter
\renewcommand{\theequation}{S\arabic{equation}}
\renewcommand{\thetable}{S\Roman{table}}
\renewcommand{\thefigure}{S\arabic{figure}}

\section{The tight-binding model}

The eight-band tight binding model with next-nearest neighbor hopping in
germanene/MoS$_2$ is given by
\begin{eqnarray}
H_{0}=\sum_{\alpha\beta}\sum_{\mu\nu\sigma}h^{\alpha\beta}_{\mu\nu}(\mathbf{k})c^\dag_{\alpha\mu\sigma}(\bm{k})c_{\beta\nu\sigma}(\bm{k}),
\end{eqnarray}
where $\alpha/\beta=A,B$ is the sublattice index and $\mu/\nu=1,2,3,4$
refer to the $s$, $p_z$, $p_x$ and $p_y$ orbitals, respectively.
$c^\dag_{\alpha\mu\sigma}$ creates a spin $\sigma$ electron in $\mu$
orbital on $\alpha$ sublattice with momentum $\bm{k}$. The nonzero
matrix elements $h^{\alpha\beta}_{\mu\nu}(\mathbf{k})$ are given by,
\begin{eqnarray}
h^{AA/BB}_{11}(\mathbf{k})&=&\epsilon_s+R^{11}(2cosk_y a_0+4cos\frac{\sqrt{3}}{2}k_x a_0cos\frac{1}{2}k_y a_0)\\
h^{AA/BB}_{22}(\mathbf{k})&=&\epsilon_{p_z}+T^{11}(2cosk_y a_0+4cos\frac{\sqrt{3}}{2}k_x a_0cos\frac{1}{2}k_y a_0)\\
h^{AA/BB}_{13/31}(\mathbf{k})&=&\pm 2i\sqrt{3}S^{12}_2cos\frac{1}{2}k_y a_0sin\frac{\sqrt{3}}{2}k_x a_0\\
h^{AA/BB}_{14/41}(\mathbf{k})&=&\pm 2iS^{12}_2sink_y a_0\pm2iS^{12}_2sin\frac{1}{2}k_y a_0cos\frac{\sqrt{3}}{2}k_x a_0\\
h^{AA/BB}_{33}(\mathbf{k})&=&\epsilon_{p_x}+2T^{22}_{11}cos(k_y a_0)+(T^{22}_{11}+3T^{22}_{22})cos\frac{\sqrt{3}}{2}k_x a_0cos\frac{1}{2}k_y a_0\\
h^{AA/BB}_{44}(\mathbf{k})&=&\epsilon_{p_y}+2T^{22}_{22}cos(k_y a_0)+(3T^{22}_{11}+T^{22}_{22})cos\frac{\sqrt{3}}{2}k_x a_0cos\frac{1}{2}k_y a_0\\
h^{AA/BB}_{34/43}(\mathbf{k})&=&\sqrt{3}(T^{22}_{11}-T^{22}_{22})sin\frac{\sqrt{3}}{2}k_x a_0sin\frac{1}{2}k_y a_0\\
h^{AB}_{11}(\mathbf{k})&=&r^{11}(e^{\frac{ik_xa_0}{\sqrt{3}}}+2e^{-\frac{ik_xa_0}{2\sqrt{3}}}cos \frac{k_y a_0}{2})\\
h^{AB}_{22}(\mathbf{k})&=&t^{11}(e^{\frac{ik_xa_0}{\sqrt{3}}}+2e^{-\frac{ik_xa_0}{2\sqrt{3}}}cos \frac{k_y a_0}{2})\\
h^{AB}_{12/21}(\mathbf{k})&=&\pm s^{11}(e^{\frac{ik_xa_0}{\sqrt{3}}}+2e^{-\frac{ik_xa_0}{2\sqrt{3}}}cos \frac{k_y a_0}{2})\\
h^{AB}_{13/31}(\mathbf{k})&=&\pm[s^{12}_1e^{\frac{ik_xa_0}{\sqrt{3}}}-s^{12}_1e^{-\frac{ik_xa_0}{2\sqrt{3}}}cos \frac{k_y a_0}{2}]   \\
h^{AB}_{14/41}(\mathbf{k})&=&\pm\sqrt{3}is^{12}_1e^{-\frac{ik_x a_0}{2\sqrt{3}}}sin \frac{k_y a_0}{2}\\
h^{AB}_{23/32}(\mathbf{k})&=&t^{12}_1e^{\frac{ik_xa_0}{\sqrt{3}}}-t^{12}_1e^{-\frac{ik_xa_0}{2\sqrt{3}}}cos \frac{k_y a_0}{2}  \\
h^{AB}_{24/42}(\mathbf{k})&=&\sqrt{3}it^{12}_1e^{-\frac{ik_x a_0}{2\sqrt{3}}}sin \frac{k_y a_0}{2} \\
h^{AB}_{33}(\mathbf{k})&=&t^{22}_{11}e^{\frac{ik_xa_0}{\sqrt{3}}}+\frac{1}{2}(t^{22}_{11}+3t^{22}_{22})e^{-\frac{ik_xa_0}{2\sqrt{3}}}cos(\frac{k_ya_0}{2})\\
h^{AB}_{34/43}(\mathbf{k})&=& -\frac{\sqrt{3}i}{2}(t^{22}_{11}-t^{22}_{22})e^{-\frac{ik_xa_0}{2\sqrt{3}}}sin(\frac{k_y a_0}{2})\\
h^{AB}_{44}(\mathbf{k})&=&t^{22}_{22}e^{\frac{ik_xa_0}{\sqrt{3}}}+\frac{1}{2}(3t^{22}_{11}+t^{22}_{22})e^{-\frac{ik_xa_0}{2\sqrt{3}}}cos(\frac{k_ya_0}{2})
\end{eqnarray}
where $a_0$ is the inplane lattice constant, $r$$(R)$ represents hopping
between $s$ and $s$ orbitals for the nearest neighbors (next nearest
neighbors), $t$$(T)$ represents hopping between $p$ and $p$ orbitals for
the NN (NNN) and $s$$(S)$ represents hopping between $s$ and $p$
orbitals for the NN (NNN). The hopping parameters in the model are
\begin{eqnarray}
&&r^{11}=V_{ss\sigma}, t^{11}=cos^2\theta V_{pp\sigma}+sin^2\theta V_{pp\pi}, s^{12}_1=sin\theta V_{sp\sigma}, s^{11}_1=cos\theta V_{sp\sigma},\\
&& t^{12}_1=-cos\theta sin\theta (V_{pp\pi}-V_{pp\sigma}), t^{22}_{11}=sin^2\theta V_{pp\sigma}+cos^2\theta V_{pp\pi}, t^{22}_{22}=V_{pp\pi},\\
&&R^{11}=V'_{ss\sigma}, S^{12}_2=V'_{sp\sigma}, T^{11}=T^{22}_{11}=V'_{pp\pi}, T^{22}_{22}=V'_{pp\sigma}.
\end{eqnarray}
The corresponding parameters (in eV) in the calculations are
\begin{eqnarray}
&&\epsilon_{s}=-7.0238, \epsilon_{p_x/p_y}=1.3088,  \epsilon_{p_z}=1.9777, \theta=111.44^\circ\\
&& V_{ss\sigma}=-1.3610 , V_{sp\sigma}=1.7288, V_{pp\sigma}=3.0701,  V_{pp\pi}=-0.6206,\\
&&V'_{ss\sigma}=-0.0729, V'_{sp\sigma}=0.3184, V'_{pp\pi}=-0.1081, V'_{pp\sigma}=0.6999.
\end{eqnarray}

\section{Triplet paring from the analysis of the Fermi velocity}

The density of states generally can be written as
$\rho(E)=\frac{\Omega}{(2\pi)^3}\sum_n\int d{\bm{k}}
\delta(E-\epsilon_{n\bm{k}})$. We rewrite $d{\bm{k}}$ as
$dS_{\|}dk_{\perp}$, where $k_{\perp}$ is along the gradient direction
of $\epsilon_{n\bm{k}}$ and $d{S_{\|}}$ is the infinitesimal area of the
iso-energy surface. Using
$d\epsilon=|\triangledown_{\bm{k}}\epsilon_{n\bm{k}}|dk_{\perp}$, we can
obtain $\rho(E)=\frac{\Omega}{(2\pi)^3}\sum_n\int d{S^E_{\|}}
\frac{1}{|\triangledown_{\bm{k}}\epsilon_{n\bm{k}}|} $. When
$|\triangledown_{\bm{k}}\epsilon_{n\bm{k}}|=0$ at some $k$ point, it is
a saddle point and called van Hove singularity point. At the Fermi
level, the density of state for the $\bm{k}$ points on surfaces is
$\rho(E_f,\bm{k}_F)\sim
\frac{1}{|\triangledown_{\bm{k}}\epsilon_{n\bm{k}}|}=
\frac{1}{|v_{F}(\bm{k}_F)|}$. The k point with smaller Fermi velocity
will contribute more to the DOS.

\begin{figure}[t]
\centerline{\includegraphics[width=1.0\columnwidth]{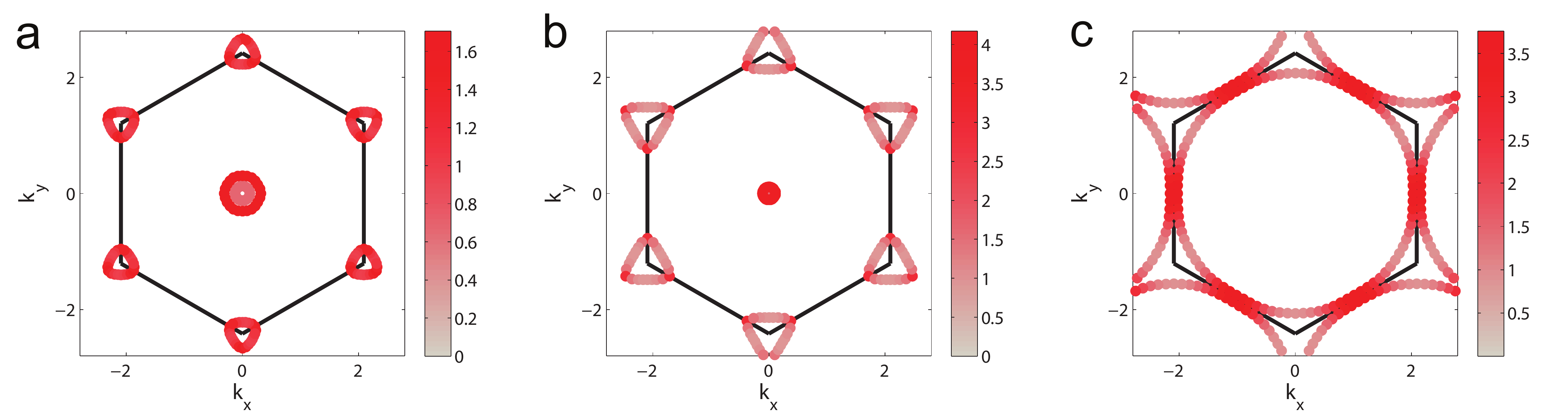}}
\caption{Distribution of the momentum resolved DOS along the Fermi
surface at half-filling (a), 0.08 electron doping (b) and at the vHs
(c).}
\label{suppfig2}
\end{figure}

For the case of germanene within the DFT Wannier model, the inverse of
the Fermi velocity is given in Fig.\ref{suppfig2} for three different
doping values. At the vHs we find that not only $k$-points around M but
also along the M-K direction are characterized by a very small Fermi
velocity, indicating that all these points give a considerable
contribution to the DOS. This is quite different, for example, from the
simple graphene model with NN hopping. The analytical RG method used by
Nandkishore et al. in Nat. Phys. {\bf 8} 158 (2012), where only the
three M points are considered, is not applicable. In germanene on
MoS$_2$, the description is very similar to the case of type-II vHs
points (the saddle points are not at the time reversal invariant
points). The saddle points contribute dominantly to DOS and enhance the
ferromagnetic fluctuations, which can promote triplet pairing. The
triplet pairing at the standard type-I vHs (saddle points at time
reversal invariant momenta) is forbidden because of the Pauli exclusion
principle.

\begin{figure}[h]
\centerline{\includegraphics[width=1.0\columnwidth]{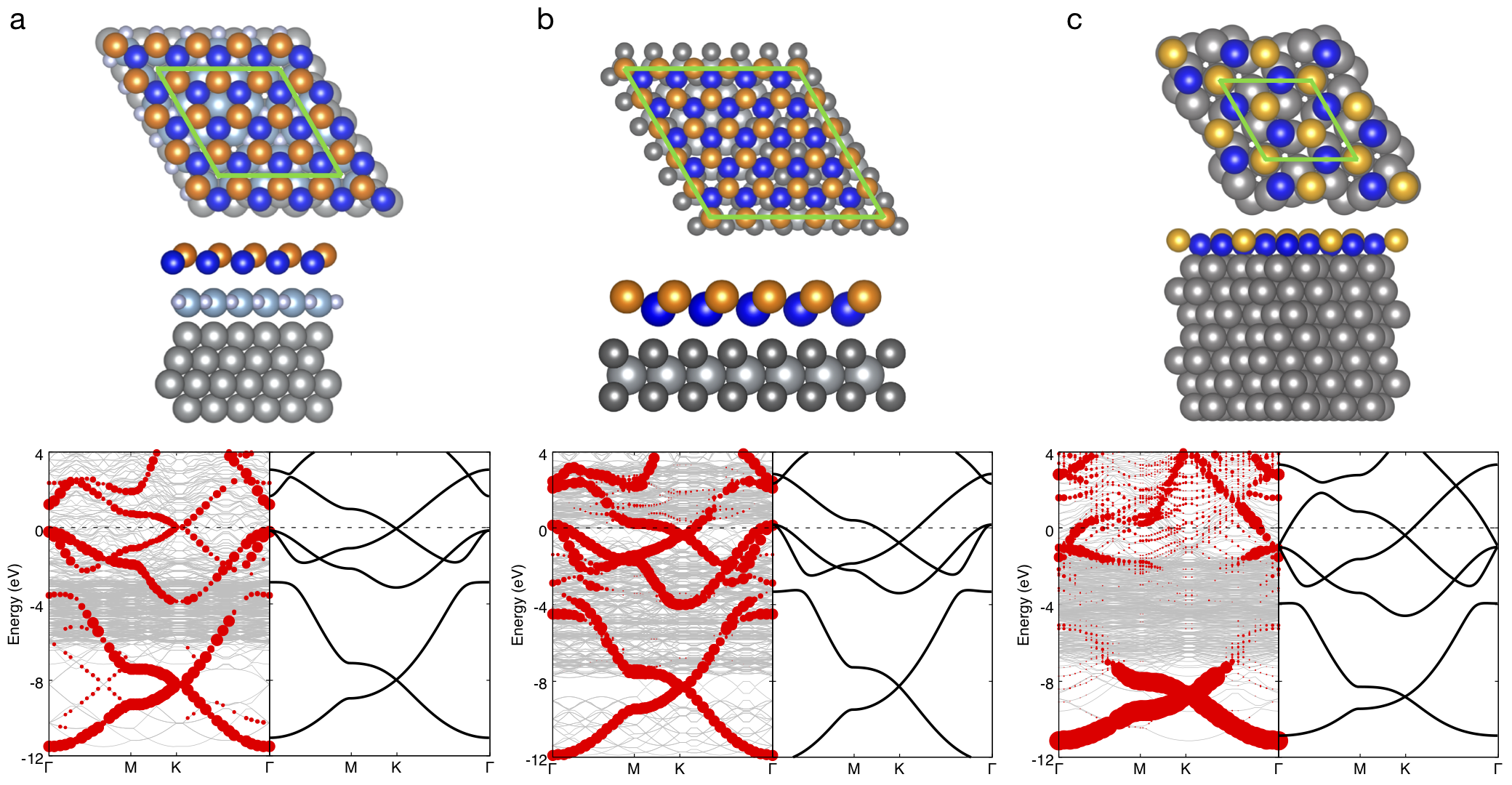}}
\caption{First-principles calculations of realistic germanene's models.
{\bf a} $3\times 3$ germanene on $4\times 4$ AlN/Ag(111), {\bf b}
$5\times 5$ germanene on $6\times 6$ MoS$_2$ and {\bf c} $\sqrt{3}\times
\sqrt{3}R(30^\circ)$ germanene on $\sqrt{7}\times \sqrt{7}R(19.1^\circ)$
Au(111). In the top and side views the orange and blue spheres refer to
protruding and low Ge atoms, respectively. The unit cell for the three
germanene's reconstructions that we considered here is in light green.
The lower panels show the DFT bandstructures for each supercell along
the high-symmetry-lines of the $1\times 1$ Brillouin zones (grey lines).
The red circles highlight the weights of the unfolded electronic states.
In black solid lines, the bandstructures of the nearest neighbour
8-bands Slater-Koster model with parameters listed in Table
(\ref{tab1}).}
\label{suppfig3}
\end{figure}

\begin{figure}[h]
\centerline{\includegraphics[width=1.0\columnwidth]{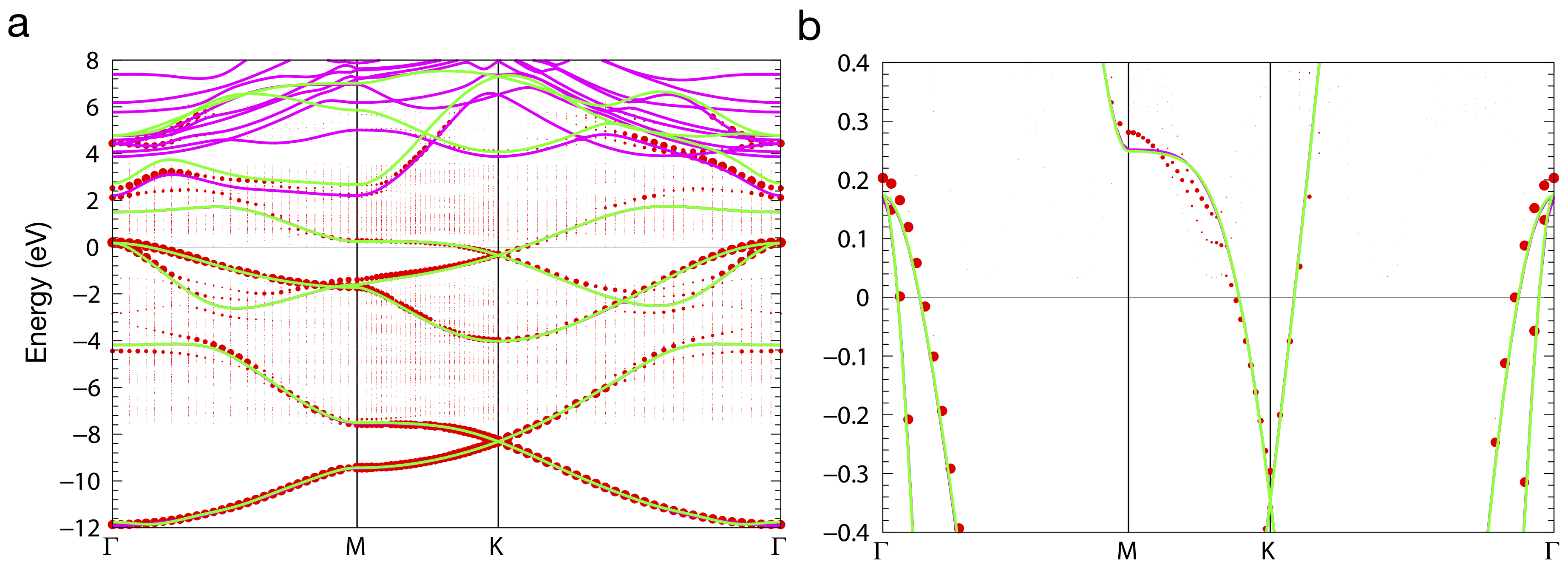}}
\caption{Large energy view and zoom of the comparison between the
bandstructures of germanene/MoS$_2$ (DFT), monolayer germanene (DFT) and
Wannier functions based Hamiltonian with $D_{3d}$ symmetry (used as the
non interacting reference for our FRG and RPA calculations).}
\label{suppfig4}
\end{figure}

\section{The influence of the buckling in germanene}

\begin{figure}[h]
\centerline{\includegraphics[width=1.0\columnwidth]{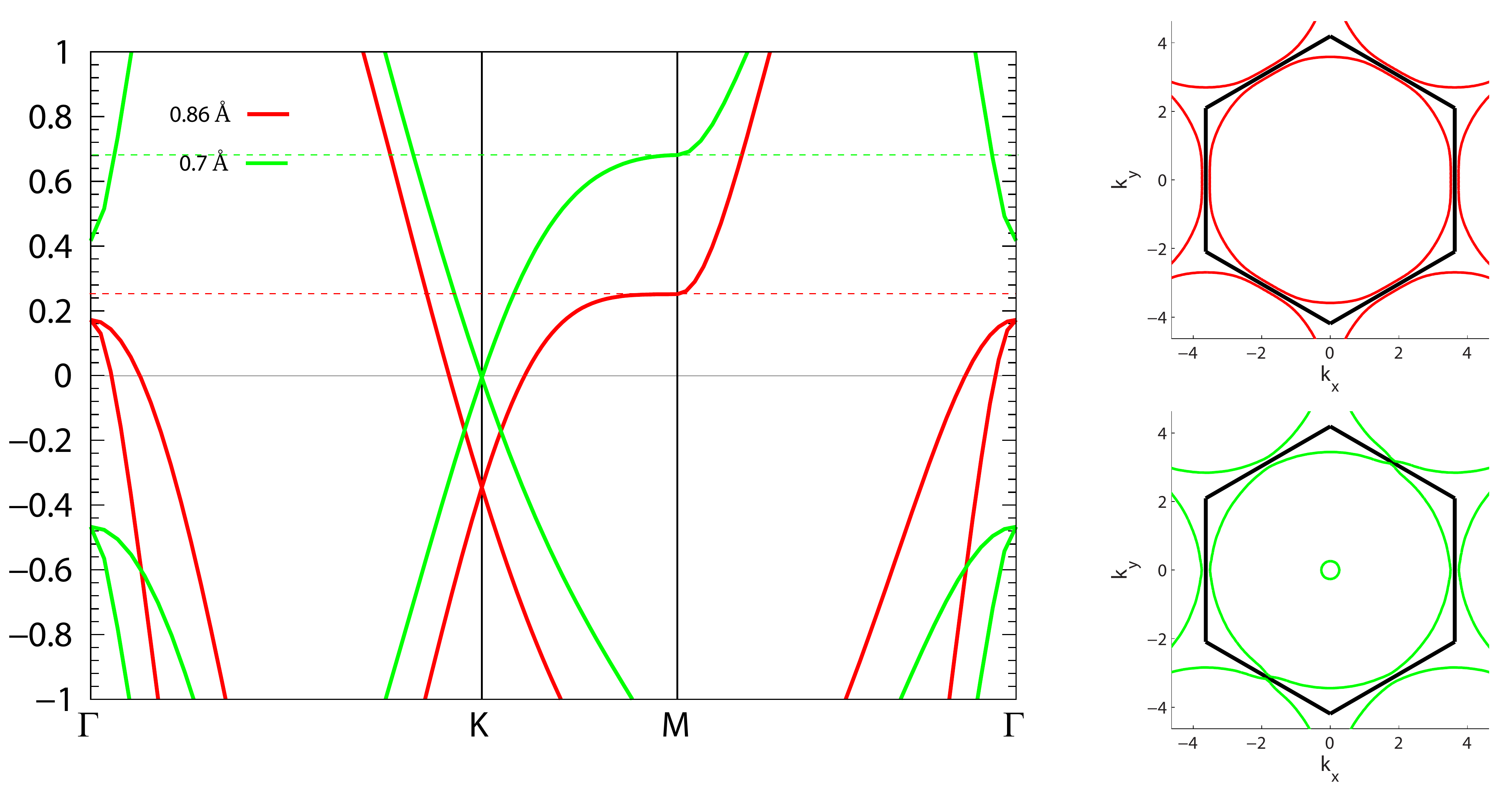}}
\caption{{\bf left} Bandstructures for germanene with different buckling
values (and relative different lattice constants). The green and red
lines denote the bandstructures for freestanding germanene and
germanene/MoS$_2$, respectively. {\bf right} Fermi surfaces at the vHs
for freestanding germanene (0.7\AA~) and germanene/MoS$_2$ (0.86\AA~).}
\label{Gebuckling}
\end{figure}

For freestanding germanene, the lattice constant is $a=4.06$\AA~ and the
buckling is about 0.7 \AA~($\theta=106^\circ$). When germanene is grown
on MoS$_2$, the lattice constant reduces to $a=3.82$\AA and the buckling
is $0.86$\AA ($\theta=111.4^\circ$)~ due to strain effects from the
substrate. In this specific case, the coupling between germanene and
MoS$_2$ is weak and the most important effect is the strain.
Fig.\ref{Gebuckling}{\bf a} shows the band structures for germanene with
different bucklings (different lattice constants), where the pink line
denotes the band structure for freestanding germanene. By increasing the
buckling, the $p_{x,y}$ band at $\Gamma$ point moves up, the VHS at M
moves down and the band around the vHs becomes flatter. When the
buckling is larger than 0.86\AA~, the $p_{x,y}$ band is above the vHs
point at half-filling and we realize a compensated semimetal phase,
where there are both electron and hole pockets at the vHs. The buckling
angle $\theta$ increases with increasing the buckling and it affects the
hopping amplitude between the A and B sublattices. The Fermi surfaces
for freestanding germanene and germanene/MoS$_2$ are shown in
Fig.\ref{Gebuckling}{\bf b},{\bf c}. We find that the Fermi surface for
the larger buckling is flatter near the M point. How to understand the
effect of the buckling on the band structures from a miscroscopic
viewpoint? By increasing $\theta$, the coupling between $p_z$ orbitals
($t^{11}$) and the coupling between $p_{x,y}$ orbitals ($t^{22}_{11}$,
$t^{22}_{22}$)  will decrease (see equations in the tight-binding model
of the Methods Section). At the $\Gamma$ point, the $p_{x,y}$ bands are
the bonding states between the A and B sublattice and therefore they
move up when the corresponding coupling decreases. The decrease of $\mid
t^{11}\mid$ leads to the downward shift of the vHs at the M point.

\end{document}